\documentclass[aps,prl,twocolumn]{revtex4-1}
\usepackage{graphicx,dcolumn,color,lipsum}
\usepackage{amsmath,amssymb,verbatim,moreverb,bm}

\begin{document}

\title{Non-equilibrium transport through a model quantum dot:
Hartree-Fock approximation and beyond}

\author{Christian Schiegg}
\affiliation{Institut f\"ur Physik, Universit\"at Augsburg, 86135 Augsburg, Germany}

\author{Michael Dzierzawa}
\affiliation{Institut f\"ur Physik, Universit\"at Augsburg, 86135 Augsburg, Germany}

\author{Ulrich Eckern}
\affiliation{Institut f\"ur Physik, Universit\"at Augsburg, 86135 Augsburg, Germany}


\begin{abstract}
The finite-temperature transport properties of the spinless interacting fermion model coupled to non-interacting leads are investigated. Employing
the unrestricted time-dependent Hartree-Fock (HF) approximation, the transmission probability and the non-linear $I$-$V$ characteristics are 
calculated, and compared with available analytical results and with numerical data obtained from a Hubbard-Stratonovich decoupling of the interaction.
In the weak interaction regime, the HF approximation reproduces the gross features of the exact $I$-$V$ characteristics but fails to account for
subtle properties like the particular power law for the reflected current in the interacting resonant level model.
\end{abstract}

\maketitle


\section{Introduction} 
Out-of-equilibrium quantum systems have received much attention in the past few decades, both experimentally
and theoretically \cite{Andergassen2010}.
A major goal is to understand the transport of charge and energy through molecular or nanoscale systems
coupled to reservoirs such that voltages and temperature gradients can be applied.
While there exist powerful numerical and analytical methods to calculate ground state and finite
temperature properties of isolated interacting quantum systems, the situation becomes much more
involved when these systems are coupled to reservoirs and driven out of equilibrium, even in the case when a stationary 
state is reached \cite{Hershfield1993}.
Due to these difficulties most of the previous studies have been restricted to single site models like the
spinless interacting resonant level model (IRLM) or the single impurity Anderson model, and the
main focus was on the zero temperature $I$-$V$ characteristics, in particular in the linear regime. 
A variety of methods, both numerical and analytical, have been applied like, e.g.,
the time-dependent density matrix renormalization group (tdDMRG) \cite{Schmitteckert2004,Feiguin2009,Muramatsu2013}, 
non-equilibrium Green's functions \cite{MeirWingreen1992,Jauho1994}, the time evolving block decimation method \cite{Linden2013,Jeckelmann2013}, 
iterated summation of the path integral \cite{Thorwart2008}, and time-dependent density functional theory (tdDFT) 
\cite{Brandbyge2002,Schenk2008,Dzierzawa2009,Schenk2011,Schmitteckert2013}. 
While in the Meir-Wingreen approach \cite{MeirWingreen1992} the problem is formally solved, for interacting models it is generally
not possible to evaluate the various Green's functions needed as input without further approximations. In the purely numerical methods
there exist severe limitations with respect to size and dimensionality of the systems that can be studied,
and even for single site models the approaches are computationally very expensive.  

In the present study we utilize the Hartree-Fock approximation in order to calculate the $I$-$V$ characteristics
of a spinless fermion model that can be seen as a generalization of the IRLM to several sites.
The obvious advantages of this method are the relatively low computational costs compared to numerically exact approaches,
and the great flexibility with respect to dimensionality, system size and geometry, finite temperature, and type and range 
of interactions. 
On the other hand, it is well known that HF calculations for isolated systems in the ground state or in thermal
equilibrium tend to overestimate the appearance of spurious broken symmetry phases. Therefore it is most important to
benchmark the results against exact solutions. 
The purpose of this study is to assess the reliability of the HF approach in the non-equilibrium setting
in comparison with available exact results.

\section{Model and methods}

\subsection{Spinless fermion model}
We consider a one-dimensional model of spinless fermions, where $N_\mathrm{C}$ central sites (the molecule) with nearest-neighbor interaction $U$
are coupled to a left and a right lead of $N_\mathrm{L}$ and $N_\mathrm{R}$ non-interacting sites. The hopping parameter $t_0$ 
is, for simplicity, chosen to be the same in the leads and within the molecule, while the coupling between the leads and the molecule
is described by a hopping parameter $t'$ and a nearest neighbor interaction $U'$. The Hamiltonian reads
\begin{equation}
\label{Hamiltonian}
\hat H = \hat H_\mathrm{L} + \hat H_\mathrm{LC} + \hat H_\mathrm{C} + \hat H_\mathrm{CR} + \hat H_\mathrm{R} \\ 
\end{equation}
with
\begin{align}
\hat H_\mathrm{L}&= - t_0 \sum_{l=1}^{a-1} (\hat c^\dagger_l \hat c^{}_{l+1} + {\text{h.c}.}) \\
\nonumber
\hat H_\mathrm{R}&= - t_0 \sum_{l=b+1}^{N-1} (\hat c^\dagger_l \hat c^{}_{l+1} + {\text{h.c}.}) \\
\nonumber
\hat H_\mathrm{C}& = \sum_{l=a+1}^{b-1} \left\{-t_0(\hat c^\dagger_{l} \hat c^{}_{l+1} + {\text{h.c}.}) 
 + U \left(\hat n_{l} -\frac{1}{2}\right)\left(\hat n_{l+1}-\frac{1}{2}\right)\right\}  \\
\nonumber
 \hat H_\mathrm{LC}& = -t'(\hat c^\dagger_{a} \hat c^{}_{a+1} + {\text{h.c.}}) 
+ U' \left(\hat n_{a} -\frac{1}{2}\right)\left(\hat n_{a+1}-\frac{1}{2}\right)\\
\nonumber
\hat H_\mathrm{CR}& = -t'(\hat c^\dagger_{b} \hat c^{}_{b+1} + {\text{h.c.}}) 
+ U' \left(\hat n_{b} -\frac{1}{2}\right)\left(\hat n_{b+1}-\frac{1}{2}\right) ,
\end{align}
where $\hat n_l = \hat c^\dagger_{l} \hat c^{}_{l}$.
The sites $a+1$ and $b$ are the first and the last sites of the molecule, respectively (see Fig.~1), 
and the system size is $N = N_\mathrm{L} + N_\mathrm{C} + N_\mathrm{R}$.
\begin{figure}
\includegraphics[width=0.45\textwidth]{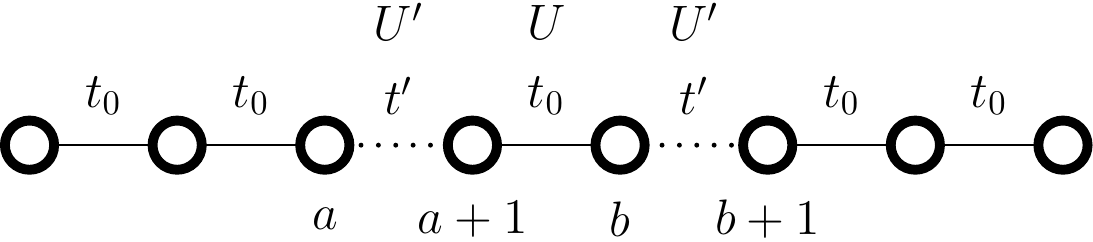}
\caption{Illustration of the model of Eq.~(\ref{Hamiltonian}) for $N_\mathrm{L} = N_\mathrm{R} = 3$ and $N_\mathrm{C} = 2$.}
\label{fig1}
\end{figure}
The charge currents between the leads and the molecule can be obtained from the equation of motion in the Heisenberg picture. Assuming that
each fermion carries the charge $e$, the charge current operator from the left lead into the molecule is given by 
\begin{equation} 
\label{leftI}
\hat I_\mathrm{L}= -e\frac{d \hat N_\mathrm{L}}{d t} = -ie[\hat H,\hat N_\mathrm{L}] = -ie t' (\hat c^\dagger_{a} \hat c^{}_{a+1} - \hat c^\dagger_{a+1} \hat c^{}_{a}) ,
\end{equation}
where $\hat N_\mathrm{L}$ is the number of particles in the left lead, and $\hbar$ is set to one.
Correspondingly, the operator of the current from the molecule into the right lead reads
\begin{equation}
\label{rightI}
\hat I_\mathrm{R} = e\frac{d \hat N_\mathrm{R}}{d t} = ie[\hat H,\hat N_\mathrm{R}] = -ie t' (\hat c^\dagger_{b} \hat c^{}_{b+1} - \hat c^\dagger_{b+1} \hat c^{}_{b}) .
\end{equation}
Due to particle number conservation, in a stationary state the expectation values of left and right currents are identical,
$\langle \hat I_\mathrm{L} \rangle =  \langle \hat I_\mathrm{R} \rangle$.

In order to probe the transport properties of a system we drive it out of equilibrium by applying
a voltage and/or a temperature gradient. There are two different schemes that have been proposed in the literature 
\cite{Caroli1971,Cini1980}
to simulate a non-equilibrium condition.
In the first setting the molecule is initially considered non-interacting and decoupled from the leads, and each of the three isolated
subsystems is assumed to be in grand canonical thermal equilibrium at its own temperature and chemical potential. 
At time $t=0$ the interaction is switched on, and
the molecule and the leads are connected by adding $\hat H_\mathrm{LC}$ and $\hat H_\mathrm{CR}$. 
The time evolution of the system is then governed by the full Hamiltonian. For sufficiently long leads one expects that
after a transient phase a stationary state arises where the current is time-independent, and 
the goal is to calculate this steady state current as function of the applied temperature and voltage bias. 

In the second scheme one considers a situation where initially the leads and the molecule are coupled and in thermal 
equilibrium at the same temperature and chemical potential. At time $t = 0$ a voltage $V$ is applied by adding different potentials $\pm eV/2$ 
to the leads. Subsequently, the time evolution is governed by the full Hamiltonian {\it including} the voltage bias. The stationary 
currents thus obtained coincide with the ones of the first quench scheme as long as the applied potential is much smaller 
than the band width of the leads \cite{Branschaedel2010}. However, for larger voltages there are significant differences in the $I$-$V$ characteristics.
While in the partitioned scheme the current approaches a finite value in the limit $V \to \infty$,
in the second scheme it goes to zero when the potential difference exceeds the band width.

Since it is not clear how to realize a well-defined temperature bias in the second setting, and since we want to retain the possibility
to treat both voltage and temperature gradients we use the first scheme for our calculations. 
The merits and shortcomings of both quench schemes have been extensively discussed in the literature 
\cite{Stefanucci2004,Branschaedel2010,Muramatsu2013}.

\subsection{Non-interacting system}
Generally, the state of a quantum mechanical system is described by the density matrix $\hat \rho$. 
In our case each subsystem $\alpha = $ L,C,R is initially in thermal equilibrium with inverse temperature $\beta_\alpha$ 
and chemical potential $\mu_\alpha$, and the density matrix reads
\begin{equation} 
\hat \rho(0) = Z^{-1} {\rm e}^{-\beta_\mathrm{L}(\hat H_\mathrm{L}-\mu_\mathrm{L}\hat N_\mathrm{L})} \
                 {\rm e}^{-\beta_\mathrm{C}(\hat H_\mathrm{C}-\mu_\mathrm{C}\hat N_\mathrm{C})}
                 {\rm e}^{-\beta_\mathrm{R}(\hat H_\mathrm{R}-\mu_\mathrm{R}\hat N_\mathrm{R})} ,
\end{equation}
where $Z$ is determined from the condition $\text{Tr}\{\hat \rho(0)\} = 1$.
The time-evolution of the density matrix is determined through
\begin{equation}
\label{rhooft}
\hat \rho(t) = {\rm e}^{-i \hat H t} \hat \rho(0) {\rm e}^{i \hat H t} .
\end{equation}
In order to calculate time-dependent expectation values of observables, like the current, one has to compute the 
equal-time Green's function defined as
\begin{equation}
G_{l m} = \langle \hat c^\dagger_{m}\hat c^{}_{l} \rangle = {\text Tr}\{\hat \rho(t) \hat c^\dagger_{m}\hat c^{}_{l} \} .
\end{equation}
For non-interacting systems, i.e., when the Hamiltonian $\hat H$ contains only bilinear combinations of the Fermi operators, 
$G$ can be obtained from the equation of motion
\begin{equation}
\label{eqofmotion}
\frac{d}{dt} G(t) = -i [H,G] ,
\end{equation}
where the matrix $H$ is defined by
\begin{equation}
\hat H = \sum_{l,m} H_{l m} \hat c^\dagger_{l}\hat c^{}_{m} .
\end{equation}
Equation~(\ref{eqofmotion}) is valid also for time-dependent but non-interacting Hamiltonians.
Expectation values of arbitrary products of Fermi operators can be expressed in terms of the Green's function $G_{l m}$ 
by applying Wick's theorem. 
Therefore, once the matrix $H$ is numerically diagonalized, it is straightforward to calculate time-dependent charge 
currents for arbitrary times. 
 
\subsection{Hartree-Fock approach}
When interactions are taken into account the task of determining the time-evolution of the density matrix becomes
much more involved. Exact solutions for interacting systems out of equilibrium are extremely rare and limited to very special
points in the parameter space of the considered model, e.g., the self dual point $U' = 2t_0$ of the IRLM \cite{Boulat2008}. 
A rather simple and physically intuitive approximation arises from the Hartree-Fock decoupling of the
interaction terms in the Hamiltonian,
\begin{align}  
\label{hafo}
\nonumber
\hat n_l \hat n_{l+1} \to &\langle\hat n_l\rangle \hat n_{l+1} + \langle \hat n_{l+1}\rangle \hat n_{l} 
- \langle \hat n_{l+1}\rangle \langle \hat n_{l}\rangle
- \langle \hat c^\dagger_{l}\hat c^{}_{l+1}\rangle \hat c^\dagger_{l+1}\hat c^{}_{l}  \\
 &
- \langle \hat c^\dagger_{l+1}\hat c^{}_{l}\rangle \hat c^\dagger_{l}\hat c^{}_{l+1} 
+ \langle \hat c^\dagger_{l+1}\hat c^{}_{l}\rangle \langle \hat c^\dagger_{l}\hat c^{}_{l+1}\rangle .
\end{align}
As a result, the time-evolution of the density matrix is governed by a non-interacting but time-dependent Hamiltonian
$\hat H_\mathrm{HF}$, and the equation of motion for the one-particle density matrix reads
\begin{equation}
\frac{d}{dt} G_\mathrm{HF}(t) = -i [H_\mathrm{HF}(t), G_\mathrm{HF}(t)] .
\end{equation}
This equation of motion can be solved numerically with arbitrary precision using the short-time propagation
\begin{equation}
G_\mathrm{HF}(t + \Delta t) \simeq {\rm e}^{-iH_\mathrm{HF}(t) \Delta t} G_\mathrm{HF}(t) {\rm e}^{iH_\mathrm{HF}(t) \Delta t} 
\end{equation}
and choosing sufficiently small time steps $\Delta t$.

So far, we have discussed the time-evolution of the system starting from a given initial state in order to
extract the transport properties from the stationary state that emerges in the course of time.
As an alternative, one can use an approach that allows one to calculate the stationary currents directly
without considering explicitly the time-evolution.
Formally this is achieved by the Meir-Wingreen approach \cite{MeirWingreen1992,Jauho1994}, 
based on the non-equilibrium Green's function formalism.
For an effectively non-interacting system like $\hat H_\mathrm{HF}$ 
their result for the $I$-$V$ relation is equivalent to the Landauer formula \cite{Landauer1970,Ness2010}.
It has been shown \cite{Brandbyge2007} that the stationary state Green's function $G^{\mathrm{HF}}$
can be expressed in terms of scattering states as
\begin{equation}
\label{brandbyge}
G^\mathrm{HF}_{l m} = \sum_{\alpha=L,R} \sum_{k} f_{\alpha}(\epsilon_k) \langle l| k\alpha\rangle \langle k\alpha|m\rangle , 
\end{equation}
where $|k\alpha\rangle$ is the eigenstate of $H_\mathrm{HF}$ that corresponds to an incoming wave 
from lead $\alpha = $ L,R with wave number $k$,
and $f_\alpha(\epsilon_k) = ({\rm e}^{\beta_\alpha(\epsilon_k - \mu_\alpha)} + 1)^{-1}$ with
$\epsilon_k = -2t_0 \cos k$ is the Fermi function that accounts for the thermal occupation of the states within each lead.
Explicitly, up to a normalization constant the plane wave scattering state originating in the left lead is given by
\begin{equation}
\label{scattering}
\langle m| kL\rangle = 
\left\{ 
\begin{array}{ccc}
{\rm e}^{ikm} + r_k\ {\rm e}^{-ikm} & \quad\text{for}\quad & m \in \mathrm{L} \\
t_k\ {\rm e}^{ikm} & \quad\text{for}\quad & m \in \mathrm{R}
\end{array}
\right. ,
\end{equation}
and $k = (2\pi/N_\mathrm{L}) \nu$ with $0 < \nu < N_\mathrm{L}/2$. Here we have assumed for simplicity that the wave 
functions in both leads are continued
periodically up to infinity, and that $N_\mathrm{L} = N_\mathrm{R}$. 
Numerically it is straightforward to connect the plane wave ansatz (\ref{scattering}) in the left and right lead by solving 
the Schr\"odinger equation for the HF Hamiltonian within the molecule.
Since $H_\mathrm{HF}$ itself depends on the equal-time Green's function, Eq.~(\ref{brandbyge}) has to be solved self-consistently.
Numerically this is achieved through iteration. Starting with some reasonable guess for $G^{\mathrm{HF}}$ one calculates $H_\mathrm{HF}$ 
using Eq.~(\ref{hafo}) and from there again $G^{\mathrm{HF}}$ via Eq.~(\ref{brandbyge}). The procedure is iterated until convergence 
is reached. From the self-consistent solution the steady state current can then be calculated using Eq.~(\ref{leftI}). 

The expression for the stationary current can also be cast into the form of the usual Landauer formula 
\begin{equation}
I = \frac{e}{h} \int d\epsilon (f_\mathrm{L}(\epsilon) - f_\mathrm{R}(\epsilon)) T(\epsilon) .
\end{equation}
where $T(\epsilon_k) = |t_k|^2$ is the transmission probability.
In contrast to the non-interacting case, here 
$T(\epsilon)$ is not only a function of the energy, but depends also on the voltage and the temperature of the leads
due to the self-consistency condition.

\subsection{Discrete Hubbard-Stratonovich transformation}

If the number of interacting sites is small, ideally if there is only interaction across a single link of neighboring sites
as it is the case for $N_\mathrm{C} = 2$ and $U' = 0$, it is possible to calculate the time-dependent density matrix $\hat \rho(t)$
exactly with moderate numerical effort.
The starting point is to write the time evolution operator $\hat U$ as a product of $M$ short-time evolution operators,
\begin{equation}
\hat U(t) = {\rm e}^{-i \hat H t} = {\prod}_{m=1}^{M}{\rm e}^{-i \hat H \Delta t} ,
\end{equation}
with $t = M \Delta t$. For small $\Delta t$ one may further split each exponential by dividing $\hat H = \hat T + \hat V $ 
into non-interacting contributions $\hat T$ and interaction terms $\hat V$. Using the symmetric Trotter breakup, one obtains     
\begin{equation}
{\rm e}^{-i \hat H \Delta t} = {\rm e}^{-i (\hat T/2 + \hat V +  \hat T/2)\Delta t} 
\simeq {\rm e}^{-i \hat T \Delta t/2} {\rm e}^{-i \hat V\Delta t} {\rm e}^{-i \hat T \Delta t/2} 
\end{equation}
with an error of ${\cal{O}}(\Delta t^3)$. Finally, each exponential containing $\hat V$ can be replaced by a sum over an Ising 
variable $s = \pm1$ using Hirsch's discrete Hubbard-Stratonovich decoupling \cite{Hirsch1983}
\begin{equation}
{\rm e}^{-i U (\hat n_l - 1/2) (\hat n_{l+1} - 1/2)\Delta t} = \frac{1}{2}{\rm e}^{-i U\Delta t/4}
\sum_{s=\pm 1} {\rm e}^{-i \alpha s (\hat n_l - \hat n_{l+1})\Delta t}
\end{equation}
with $\alpha\Delta t = \arccos \left({\rm e}^{i U\Delta t/2}\right)$. Applying this transformation to both exponentials in Eq.~(\ref{rhooft})
one may express $\hat \rho(t)$ as the summation over the $4^M$ configurations of $2M$ Ising variables, each of them representing the 
time evolution of a non-interacting system in a time-dependent potential. For not too long leads and a number of time slices 
$M \leq 10$
one can compute these sums numerically without introducing any further source of error.
The discrete Hubbard-Stratonovich decoupling has previously been used in combination with an iterative summation scheme in order to calculate
transport properties of the single impurity Anderson model \cite{Thorwart2008}.
We will use the Hubbard-Stratonovich method later to benchmark the results of the Hartree-Fock approximation. 
\begin{figure}
\includegraphics[width=0.45\textwidth]{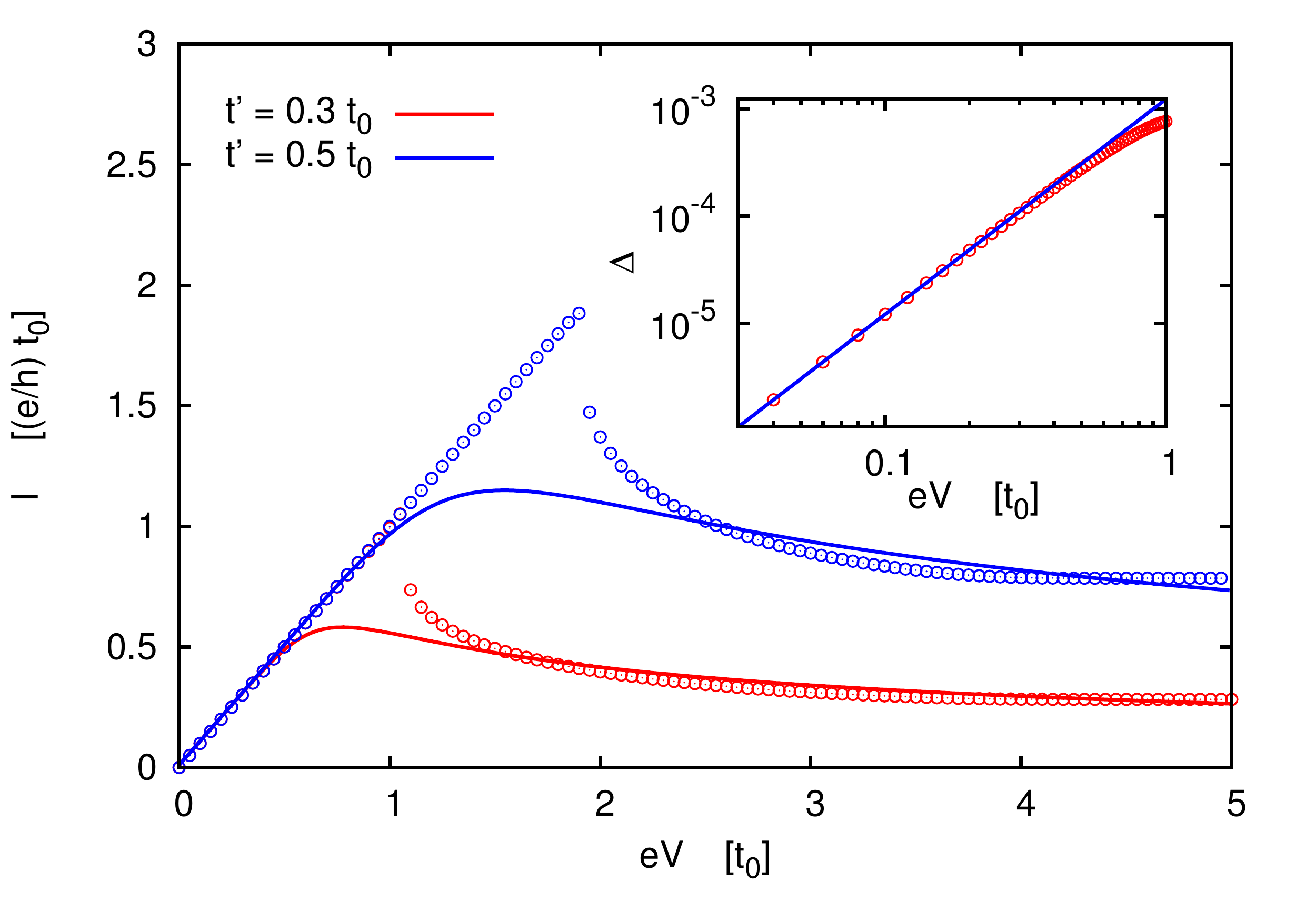}
\caption{ 
Current voltage characteristics of the IRLM for $U' = 2t_0$ and 
$t'/t_0 = 0.3, 0.5$. Symbols: HF approximation, full curves: exact results from Eq.~(\ref{hyper}).
The inset shows (for $t' = 0.5 t_0$) the relative deviation $\Delta = (I_0 - I)/I_0$ from perfect conduction, 
$I_0 = G_0 V$, $G_0 = e^2/h$, on a double logarithmic
scale. Symbols: HF data, solid line: $\Delta \propto V^2$.
}
\label{fig2}
\end{figure}
\begin{figure}
\includegraphics[width=0.45\textwidth]{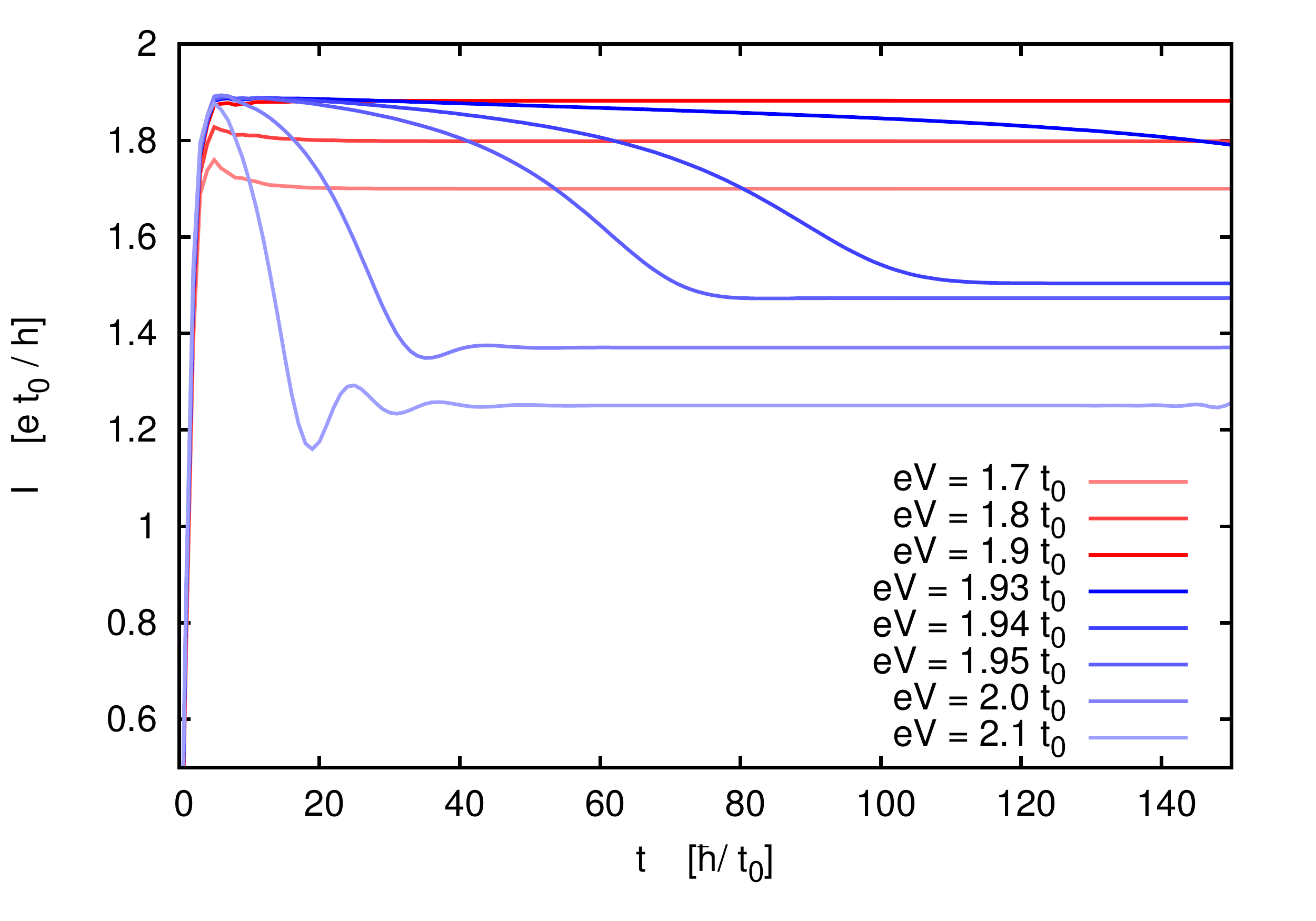}
\caption{ 
Current in HF approximation as function of time for the IRLM for $U' = 2t_0$, 
$t' = 0.5 t_0$, and several voltages close to the sharp transition at $V_c \approx 1.9 t_0/e$ 
in the $I$-$V$ curve in Fig.~\ref{fig2}.
}
\label{fig3}
\end{figure}
\section{Results}

In the following we present numerical results for the model defined in Eq.~(\ref{Hamiltonian}). We restrict ourselves 
to the cases $N_\mathrm{C} = 1$ which corresponds
to the IRLM, and $N_\mathrm{C} = 2$ which we will refer to as two-site model, for brevity. In the IRLM $U'$ is the only interaction parameter, 
whereas in the two-site model we set $U' = 0$ and vary the interaction $U$ between the two atoms of the molecule. 

The interaction dependence of the linear conductance for a two-site model
similar to ours  was studied in \cite{Moliner2011} using DMRG. In contrast to our work, in that paper the 
interaction between molecule and leads was varied, while the interaction on the molecule
was kept fixed.

Unless stated otherwise, in all calculations the length of the leads is $N_\mathrm{L} = N_\mathrm{R} = 100$, 
the inverse temperature is $\beta_{\mathrm L} = \beta_{\mathrm R} = 20/t_0$,
and the overall chemical potential of the unbiased system is $\mu = 0$ which corresponds to half-filled band in the leads.
The time-dependent current $I$ is calculated as the mean value of left and right currents if they are different.
The currents obtained from the plateau value reached in the time evolution are identical to the steady state currents calculated from the self-consistent
HF approach within a relative error typically of the order of $10^{-5}$, with the exception of the cases where hysteresis
occurs in the self-consistent solution.
\begin{figure}
\includegraphics[width=0.45\textwidth]{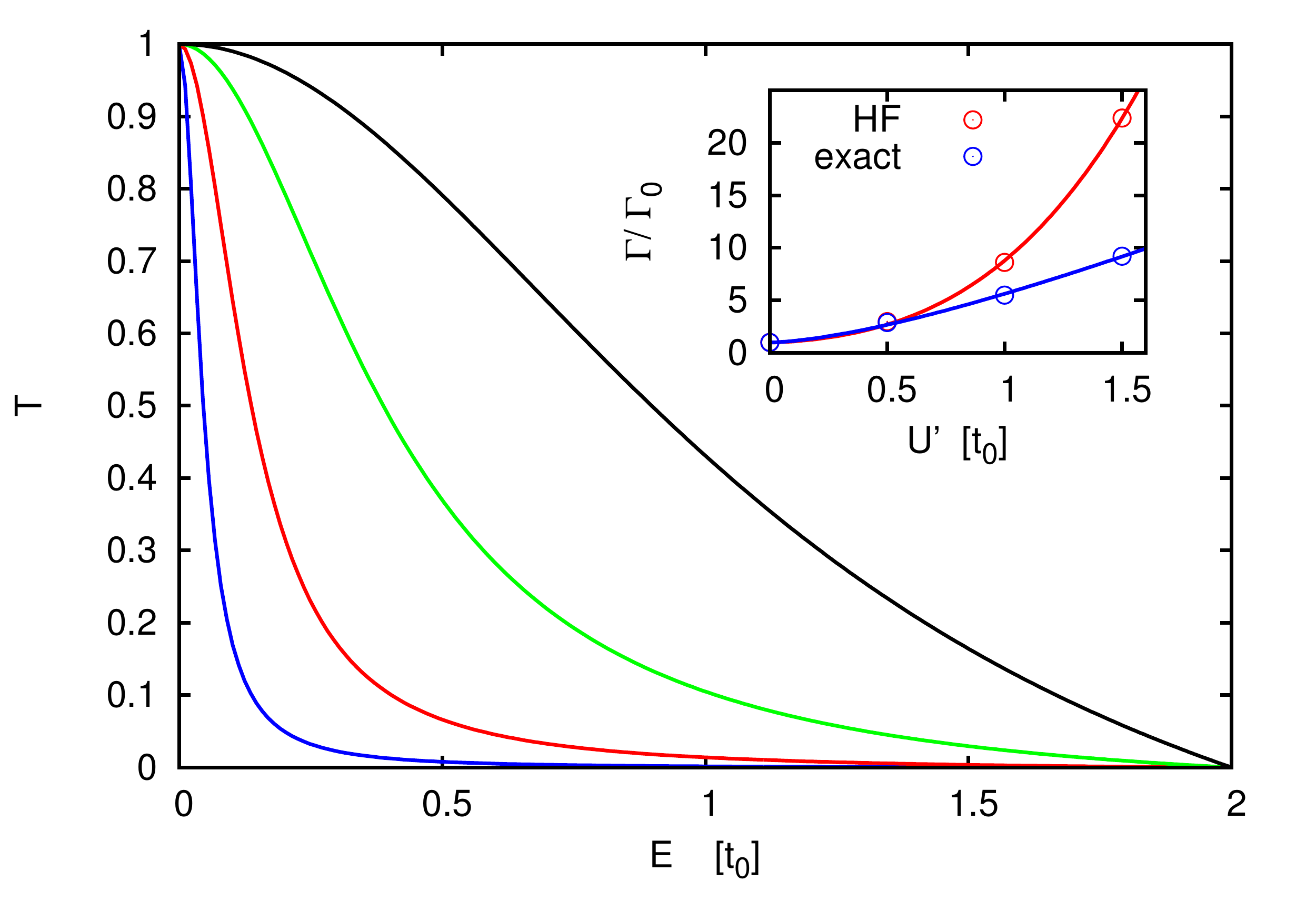}
\caption{HF transmission coefficient of the IRLM at zero voltage 
$t' = 0.15 t_0$ and $U'/t_0 = 0,0.5,1,1.5$ (from bottom to top). 
The inset shows the broadening $\Gamma/\Gamma_0$
of the peak as function of $U'$ compared to the broadening of the exact
spectral function of the zero temperature IRLM \cite{Schmitteckert2014} for the same hopping parameter.
The solid curves are polynomial interpolations as guide to eye.
}
\label{fig4}
\end{figure}
\subsection{$N_\mathrm{C} = 1$: Interacting resonant level model}

In the case where the molecule consists of a single site the Hamiltonian (\ref{Hamiltonian}) is identical to the IRLM with interaction parameter
$U'$. The $I$-$V$ characteristics of this model for the special value $U' = 2t_0$ 
is known analytically \cite{Boulat2008}, and is given in closed form by \cite{Schmitteckert2011}
\begin{equation} \label{hyper}
    I(V) = \frac{e^2 V }{2\pi \hbar } 
    \;{}_3F_2 \left[ \left\{ \frac{1}{4},\frac{3}{4}, 1 \right\}, 
                     \left\{ \frac{5}{6},\frac{7}{6}     \right\}, - \left(\frac{V}{V_{\mathrm c}}\right)^6   \right],
\end{equation}
where $\;{}_3F_2$ is a hypergeometric function, and $e V_{\mathrm c}/t_0 = r (t'/t_0)^{4/3}$.
The prefactor $r \approx 3.2$ is the lattice regularization of the corresponding field theory.
In Fig.~2 we compare the $I$-$V$ characteristics obtained within the HF approximation with the exact analytical result of Eq.~(\ref{hyper})
for hopping parameters $t' = 0.3 t_0$ and $t' = 0.5 t_0$.
For small voltages there is a linear relation between current and voltage
due to the fact that at half filling the Fermi level is exactly at the resonance, and therefore the conductance is identical to
the conductance quantum $G_0 = e^2/h$ of spinless fermions. 
Expanding Eq.~(\ref{hyper}) to leading order in the voltage yields
\begin{equation} \label{expansion}
    I(V) \approx G_0 V \left(1 - \frac{27}{140}\left(\frac{V}{V_{\mathrm c}}\right)^6\right) .
\end{equation}    
For comparison, the relative deviation $\Delta$ of the HF currents from perfect conduction is shown in the inset of Fig.~2.
A linear fit of the data on a double logarithmic scale yields $\Delta \propto V^2$, just like for the non-interacting system,  
in contrast to the nontrivial analytical result $\Delta \propto V^6$.
Perturbative calculations \cite{Boulat2014} for the IRLM away from the self-dual point $U' = 2t_0$ indicate that there exists
a nonvanishing contribution $\Delta \propto V^2$ to the backscattered nonlinear conductance although with 
a reduced weight compared to the noninteracting system. 

For larger voltages, the exact currents reach a maximum and then decrease slowly but steadily
with negative differential conductance.
The HF currents, on the other hand, increase up to somewhat larger voltages. Then a sudden drop occurs followed by a range 
of negative differential conductance quite close to the analytical result. 
For voltages beyond the band width, $eV > 4 t_0$, the HF current remains constant while the exact one continues to decrease.
The location of the jump is independent of the method how the
HF currents are calculated. In particular there is no sign of hysteresis, i.e., solving the self-consistency equations for voltages increasing
by small steps by taking the converged result of the previous voltage as starting point for the next iteration 
leads to the same curves as for incrementally decreasing voltages.
The discontinuous behavior, which is obviously an artifact of the 
HF approximation, can also be observed in
the time-evolution of the current displayed in Fig.~3 for $t' = 0.5 t_0$ and several voltages close to the jump. While for voltages
below the transition (red curves) the stationary state is reached quite fast and the plateau values increase with voltage, there is a quite
strong reduction of the stationary current within a very small range of voltages, and it takes much longer for the system to
reach the non-equilibrium steady state.

Figure 4 shows the HF transmission coefficient $T(E)$ of the unbiased (zero voltage) IRLM 
which is closely related to the local spectral function $A(E)$. At half filling $A(E)\propto {\cal{N}}(E) T(E)$, 
with the density of states
\begin{equation} \label{dos}
{\cal{N}}(E) =  \frac{1}{\pi}  \frac{1}{\sqrt{4t_0^2 - E^2}} .
\end{equation}   
The most striking feature is a pronounced broadening of the central peak with increasing interaction. 
The zero temperature spectral function of the IRLM has been calculated numerically 
using the Chebyshev expansion \cite{Schmitteckert2014}, and the broadening of the central peak has been obtained
by fitting the data to a Lorentzian,
\begin{equation} \label{lorentz}
A(E) = \frac{1}{\pi} \frac{\Gamma}{E^2 + \Gamma^2} .
\end{equation} 
In the inset of Fig.~4 we compare our data with the 
broadening calculated from the exact spectral function.
Up to $U \approx 0.5 t_0$ the HF results agree reasonably well with the exact ones but for larger interactions there
is a quite strong discrepancy indicating the limitations of the HF approximation.
%
%
%
%
%
%
\begin{figure}
\includegraphics[width=0.45\textwidth]{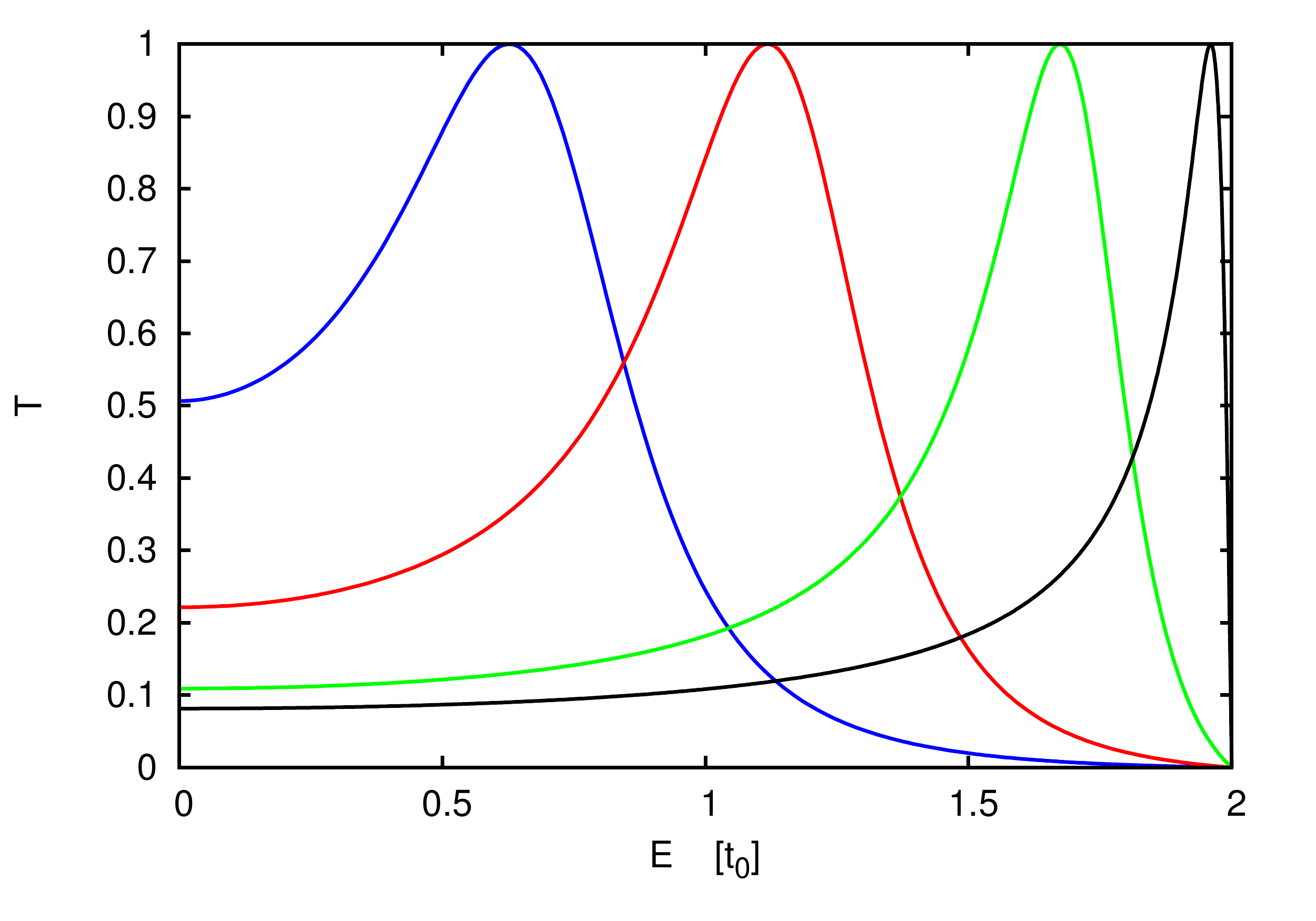}
\caption{HF transmission coefficient of the two-site model at zero voltage 
for $t' = 0.5 t_0$ and $U'/t_0 = -1,0,1,1.5$ (from left to right). 
}
\label{fig5}
\end{figure}
\subsection{$N_\mathrm{C} = 2$: Two-site model}
The transport properties of the model (\ref{Hamiltonian}) in the case $N_\mathrm{C} = 2$ are different from those of the IRLM in many respects
due to the fact that now the Fermi level lies exactly between the two transmission resonance peaks of the non-interacting system.
Therefore, the inclusion of interaction may not only broaden or shift these peaks but also modify the transmission at small energies and
thus strongly influence the conductance. 

Figure 5 shows the HF transmission coefficient of the unbiased two-site model for several values of the interaction $U$.
While for attractive interaction, $U = -t_0$, the resonance peak is shifted to smaller energy values and slightly broadened
with respect to the non-interacting one, the opposite is the case for repulsive interaction. Correspondingly, the
transmission at zero energy is reduced with increasing $U$ which is expected to result in a decreased conductance.
\begin{figure}
\includegraphics[width=0.45\textwidth]{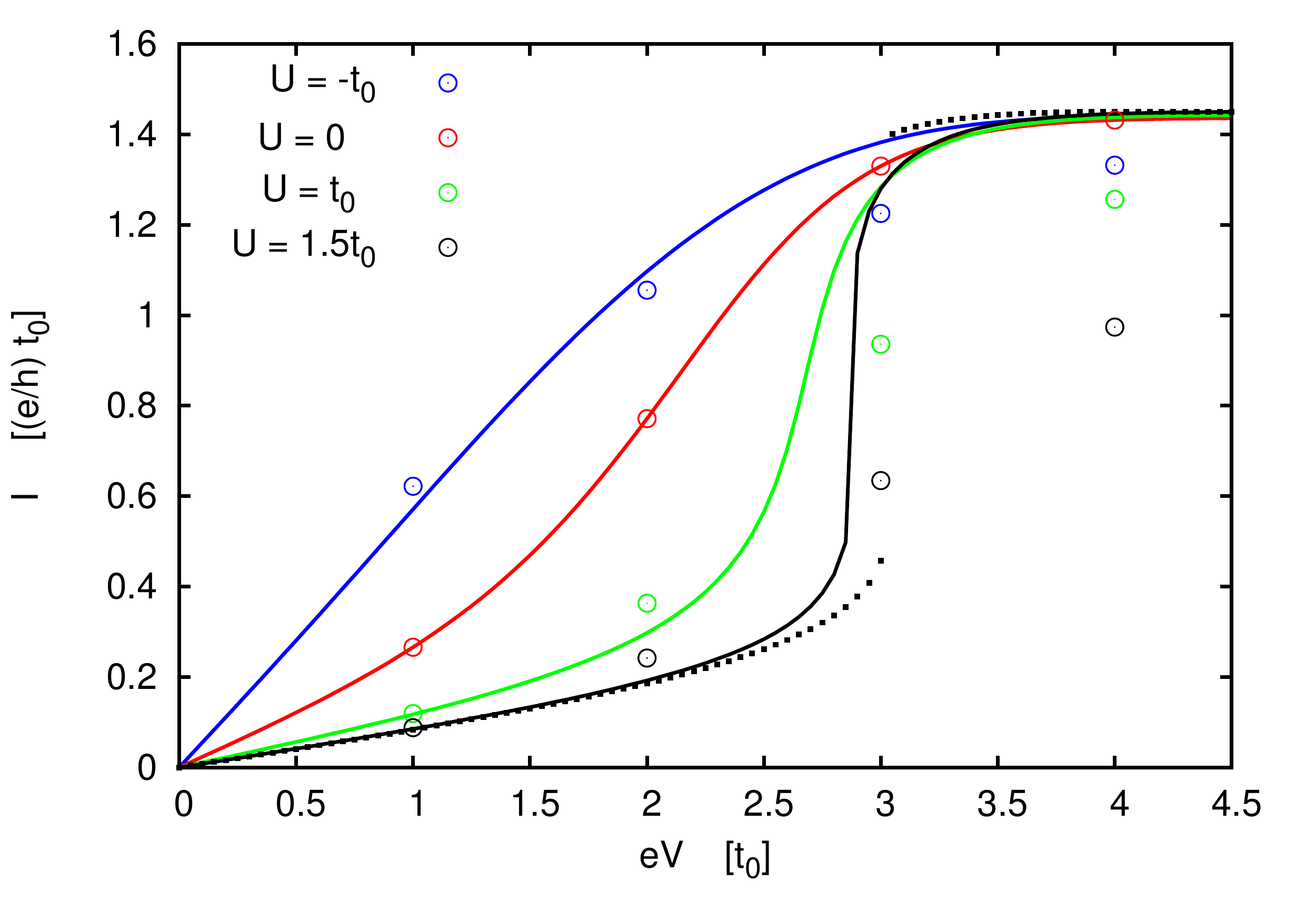}
\caption{ 
Current voltage characteristics of the two-site model for 
$t' = 0.5 t_0$, $\beta = 8/t_0$, and several values of $U$.
The curves are HF data, the symbols are obtained from the Hubbard Stratonovich approach for shorter leads with $N_\mathrm{L} = N_\mathrm{R} = 20$,
as illustrated in Fig.~7. The size of the symbols is larger than the estimated error. The dotted curve is the HF current for $U = 1.5 t_0$
and $\beta = 50/t_0$.
}
\label{fig6}
\end{figure}
\begin{figure}
\includegraphics[width=0.45\textwidth]{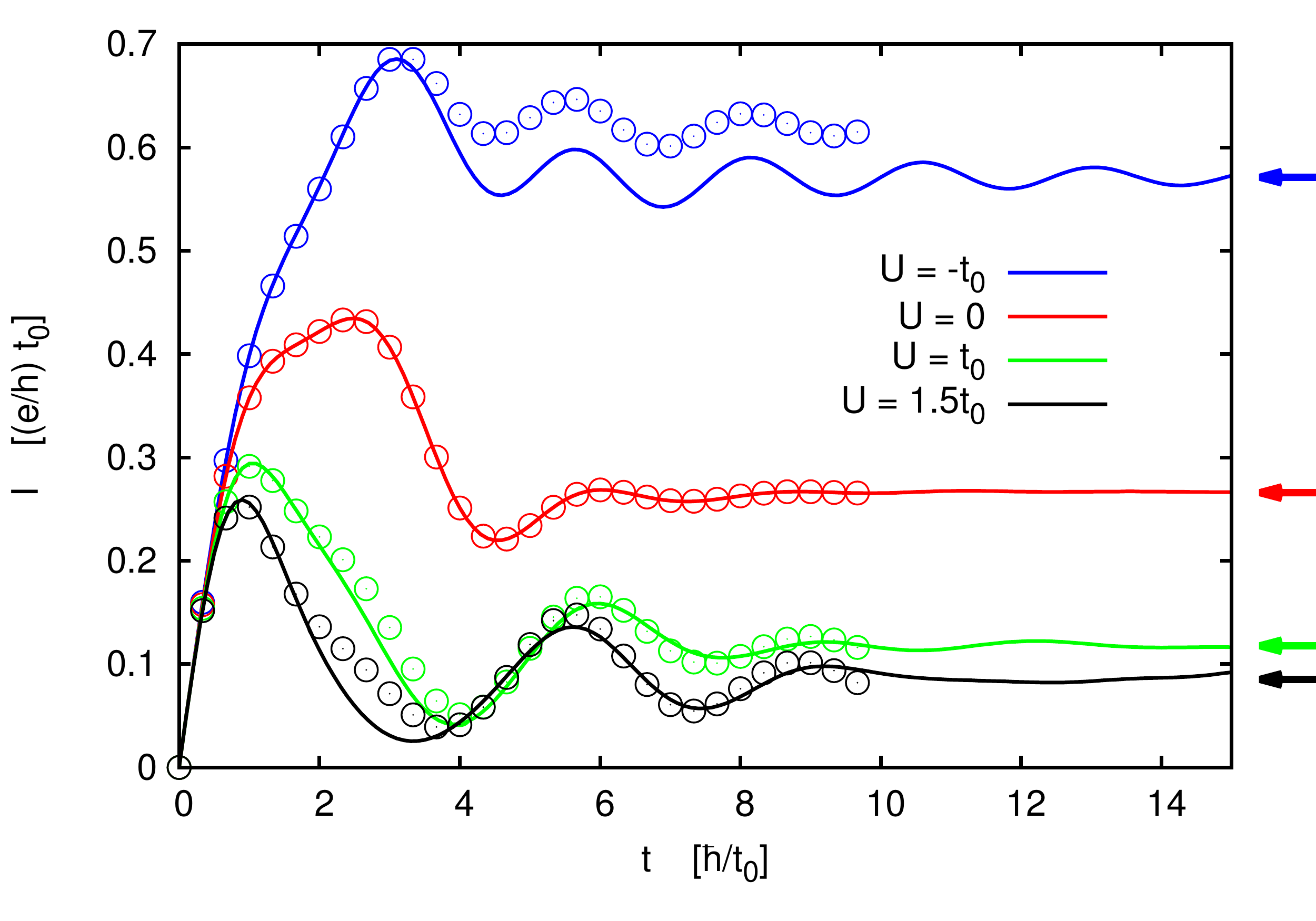}
\includegraphics[width=0.45\textwidth]{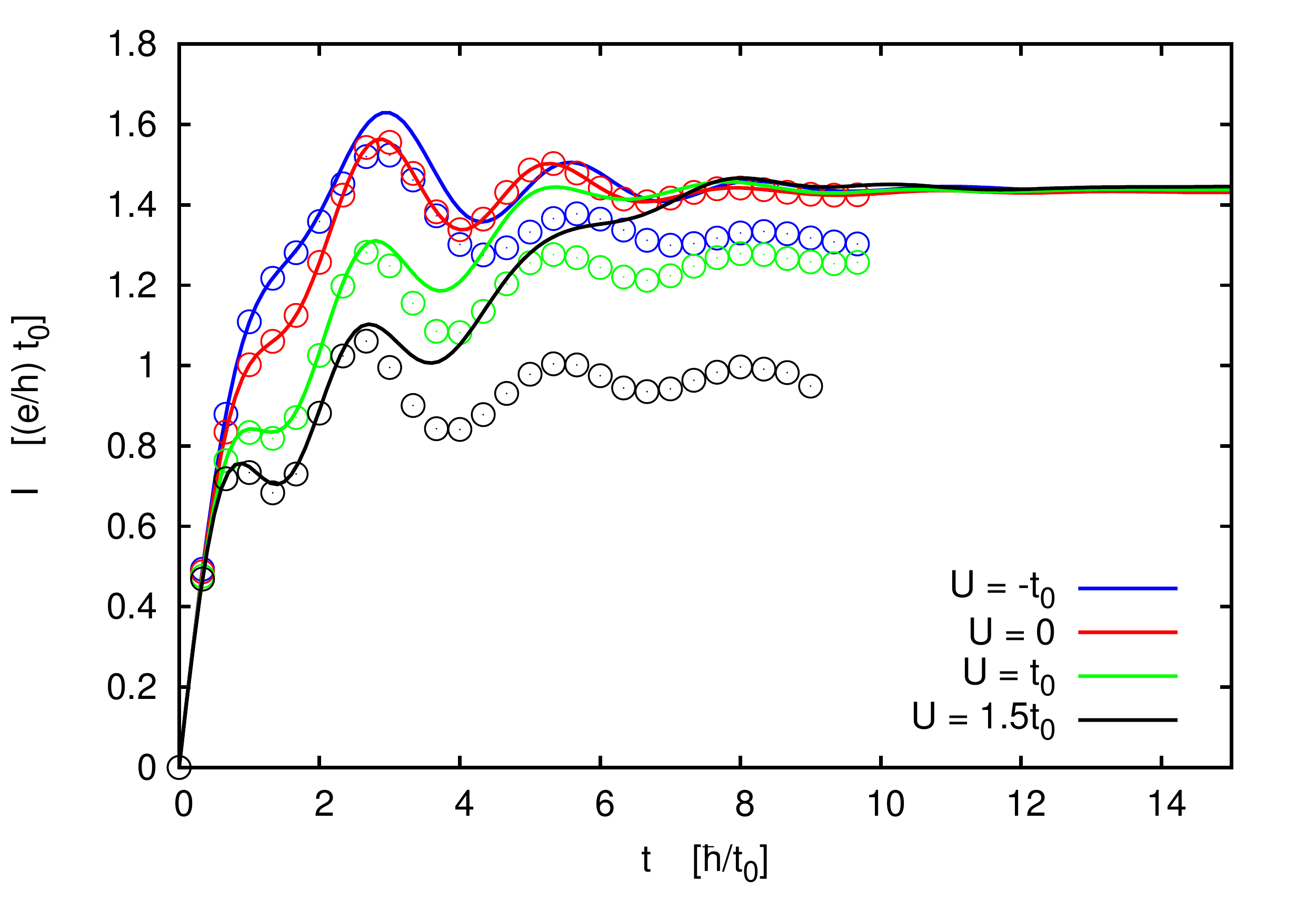}
\caption{Time-dependent currents for the two-site model with $N_\mathrm{L} = N_\mathrm{R} = 20$, 
$\beta = 8/t_0$, $t' = 0.5 t_0$, and several values of the interaction strength.
The symbols are the exact currents from the Hubbard Stratonovich approach for
$M = 9$ time slices, the full curves are HF data. 
The arrows indicate the HF stationary state currents for the same parameters 
in the limit of infinitely long leads.
Top panel: $V = t_0/e$, bottom panel: $V = 4 t_0/e$.
}
\label{fig7}
\end{figure}
This expectation is confirmed in the $I$-$V$ diagram (Fig.~6) where the linear conductance is largest for 
attractive interaction $U = -t_0$, and decreases with increasing values of $U$.
Remarkably, at higher voltages
there is a strong and for $U \ge 1.5 t_0$ even discontinuous increase of the current such that
all curves nearly collapse in the large voltage regime. 

In order to check if the strong increase of the currents in the $I$-$V$ characteristics is an artifact of the HF approximation, we have 
calculated the exact time-dependent currents using the Hubbard-Stratonovich approach described in section 2.4 
for shorter leads ($N_\mathrm{L} = N_\mathrm{R} = 20$). The number of Trotter time slices $M = 9$ is chosen such that discretization errors in the data shown in the figure 
are much smaller than the size of the symbols.
To avoid finite size effects due to the discrete spectra of the
leads, we have used a somewhat smaller inverse temperature, $\beta = 8/t_0$, for the data shown in Fig.~6 and in Fig.~7.
To get an idea about the influence of finite temperatures we also show the HF current for $U = 1.5 t_0$ and $\beta = 50/t_0$
as dotted curve in Fig.~6. The deviation from the $\beta = 8/t_0$ result is negligible with the exception of the voltage region close
to the jump.
For small voltages, $V \leq 2t_0/e$, the HF currents agree reasonably well with the exact ones displayed as open symbols of the same color,
whereas for larger voltages they are completely off. To elucidate this behavior, the time-dependent HF currents are compared
with the exact ones in Fig.~7 for two different voltages, $V = t_0/e$ and $V = 4t_0/e$. 
For the smaller voltage (upper panel), the HF currents for repulsive interaction nearly coincides with the exact
currents and deviate only slightly for repulsive interaction, $U = - t_0$. Note that the stationary currents for much longer leads indicated
by arrows on the right axis are nearly identical to what one would obtain by averaging over the small oscillations observed 
in the HF data for $N_\mathrm{L} = 20$. 

In the case of large voltage (lower panel), the HF currents resemble the exact ones only for short times during the transient phase
of the time evolution but converge all to the same current plateau of the non-interacting system. The exact currents on the other hand
approach different stationary values depending on the interaction strength $U$.

It is therefore clear that the collapse of the different curves at large voltages in the $I$-$V$ diagram of Fig.~6 is an artifact of
the HF approach and indicates a fundamental failure of the approximation in this parameter range.
This observation is in contrast to what we found for the IRLM where the HF data were in reasonable agreement with the analytical results
both for small and for large voltages, see Fig.~2. 

Interestingly, the discontinuity of the currents appearing in the $I$-$V$ diagram for sufficiently strong interaction and low temperature
can be associated with the existence of multiple solutions of the self-consistent HF equation.
In Fig.~8 the currents obtained from the iterative solution of the HF equation and the stationary current taken from the time-evolution
are shown in a small range of voltages close to the jump. Performing the iteration for voltages that increase by small steps one
obtains an $I$-$V$-curve that jumps from a low to a high current branch at a voltage of $V_{c1} \approx 3.03 t_0/e$ while in the opposite direction
the jump from high to low currents occurs at $V_{c2} \approx 2.77 t_0/e$. The true transition voltage obtained from the time-evolution 
lies in between at $V_c \approx 3 t_0/e$. 
The existence of multiple solutions in the $I$-$V$ characteristics  within the HF approximation and within the adiabatic local density 
approximation of density functional theory has recently been discussed for a model with Hubbard-type interaction \cite{Gross2012}. 

\begin{figure}
\includegraphics[width=0.45\textwidth]{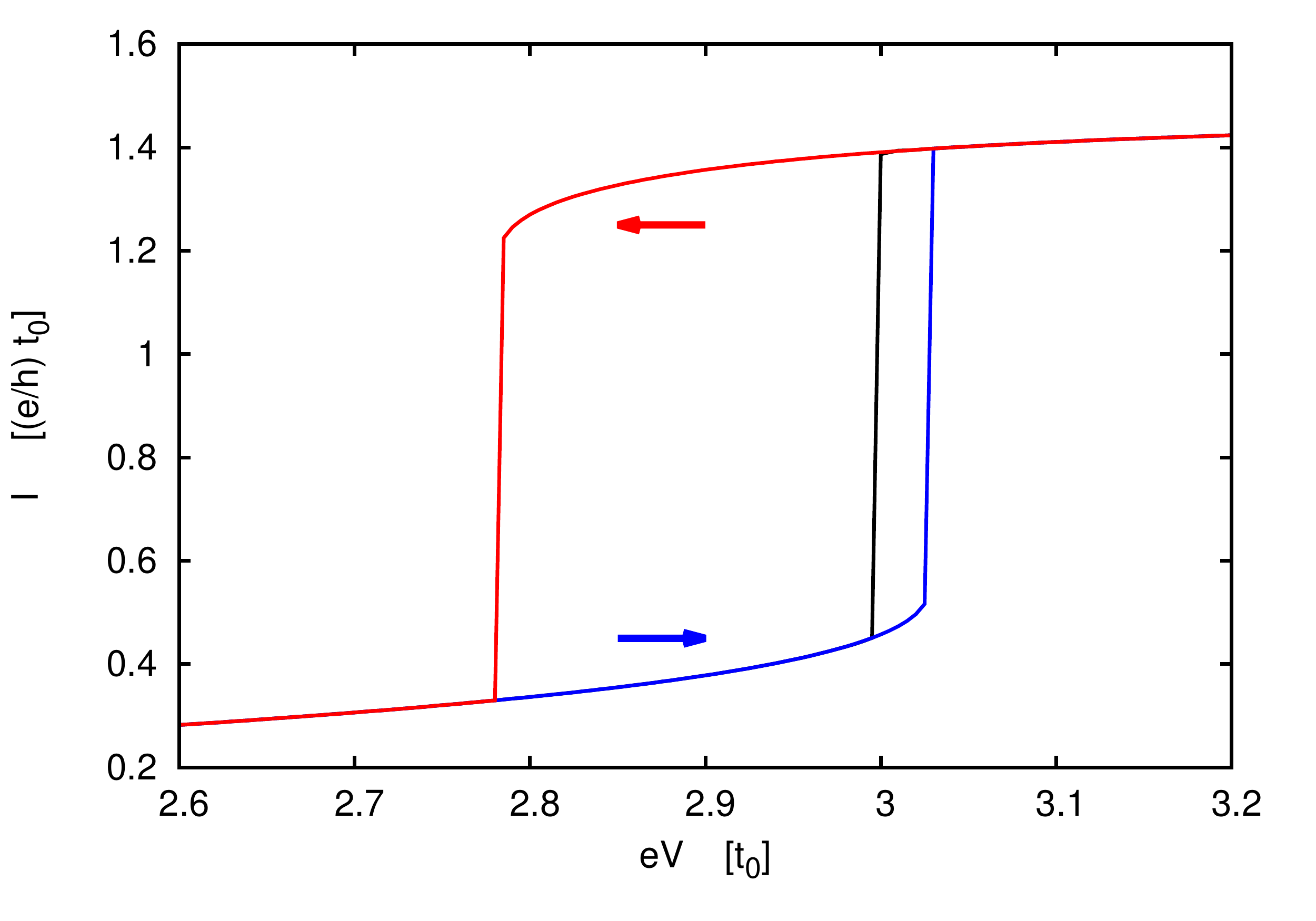}
\caption{Hysteresis in the current voltage characteristics of the two-site model for $U = 1.5t_0$, 
$t' = 0.5 t_0$, and $\beta = 50/t_0$. The blue (red) curve is obtained from the iterative solution of the HF equation (\ref{brandbyge})
for adiabatically increasing (decreasing) voltage, as indicated by the arrows. The black curve
corresponds to the stationary current taken from the time evolution.}
\label{fig8}
\end{figure}

\section{Conclusion}
The time-dependent HF approximation is a computationally cheap and versatile approach to calculate the $I$-$V$ characteristics of
weakly correlated systems at finite temperatures. The time-evolution of the currents until a plateau value is reached as well as an iterative
solution of the self-consistent HF equations for the stationary state yield identical results with comparable numerical effort.
However, the self-consistent approach sometimes allows for multiple solutions which leads to hysteretic behavior when the voltage is
varied adiabatically. This ambiguity can be avoided using the stationary current obtained from the time-evolution approach.
For a model of interacting spinless fermions, the HF data agree well with
available exact results, with the exception of the large voltage regime of the two-site model where a spurious discontinuous
transition is observed within the HF approximation. 
It is straightforward to generalize the HF approach in many respects. In addition to the charge currents also energy or heat currents can be
calculated which is of particular interest when besides the voltage there is also a temperature gradient. Furthermore, dynamical
properties, e.g., the response to a time-dependent gate voltage, can be studied without significant additional effort.

\section*{Acknowledgments}
This work has been supported by the Deutsche Forschungsgemeinschaft through TRR80.
We would like to thank Peter Schmitteckert for helpful discussions.

\end{document}